\documentclass[twocolumn,epjc3,final]{svjour3}
\RequirePackage[T1]{fontenc}

\smartqed  

\RequirePackage{graphicx}
\RequirePackage{mathptmx}         
\RequirePackage{flushend}
\RequirePackage[numbers,sort&compress]{natbib}

\usepackage{amsmath, txfonts,multirow}
\usepackage{graphicx,bm,booktabs}
\usepackage{color}
\usepackage{multirow}
\usepackage[bookmarks=true,bookmarksopen=false,plainpages=false,breaklinks=true,
   bookmarksnumbered=true,hypertexnames=false,
   filecolor=blue,urlcolor=three,menucolor=three,
   linkcolor=three,citecolor=blueone, colorlinks,
   anchorcolor=blue,runcolor=pink,frenchlinks=red
   pdfstartview=FitH,pdftitle=title,%
   pdfauthor=author]{hyperref}
   
\allowdisplaybreaks[4]

\def\tdots{\cdot\cdot\cdot}
\definecolor{cover}{rgb}{0.77,0.87,0.88}
\definecolor{blueone}{rgb}{0.1,0.1,.7}
\definecolor{citec}{rgb}{0.14,0.47,0.09}
\definecolor{two}{rgb}{0.0,0.5,0.}
\definecolor{three}{rgb}{.5,.1,0.15}

\begin{document}

\title{Hidden and doubly heavy molecular states from interactions $D^{(*)}_{(s)}{\bar{D}}^{(*)}_{s}$/$B^{(*)}_{(s)}{\bar{B}}^{(*)}_{s}$ and ${D}^{(*)}_{(s)}D_{s}^{(*)}$/${B}^{(*)}_{(s)}B_{s}^{(*)}$}
\author{Zuo-Ming Ding\thanksref{addr1}, Han-Yu Jiang\thanksref{addr1}, Dan Song\thanksref{addr1}, Jun He\thanksref{addr1,addr2,e1}}
\thankstext{e1}{Corresponding author: junhe@njnu.edu.cn}
\institute{School of Physics and Technology, Nanjing Normal University, Nanjing 210097, China\label{addr1}
\and
Lanzhou Center for Theoretical Physics, Lanzhou University, Lanzhou 730000, China\label{addr2}
}
\date{Received: date / Revised version: date}

\maketitle
\begin{abstract}
In this work, we perform a systematical investigation about the  possible hidden and doubly heavy molecular states with open  and hidden strangeness from interactions of $D^{(*)}{\bar{D}}^{(*)}_{s}$/$B^{(*)}{\bar{B}}^{(*)}_{s}$,  ${D}^{(*)}_{s}{\bar{D}}^{(*)}_{s}$/${{B}}^{(*)}_{s}{\bar{B}}^{(*)}_{s}$, ${D}^{(*)}D_{s}^{(*)}$/${B}^{(*)}B_{s}^{(*)}$, and $D_{s}^{(*)}D_{s}^{(*)}$/$B_{s}^{(*)}B_{s}^{(*)}$ in a quasipotential Bethe-Salpeter equation approach. The interactions of the systems considered are described within the one-boson-exchange model, which includes exchanges of  light mesons and $J/\psi/\Upsilon$ meson. Possible molecular states are searched for as  poles of scattering amplitudes of the interactions considered. 
The results suggest that  recently observed $Z_{cs}(3985)$ can be assigned as a molecular state of $D^*\bar{D}_s+D\bar{D}^*_s$, which is  a partner of $Z_c(3900)$ state as a $D\bar{D}^*$ molecular state. 
The calculation also favors  the existence of hidden heavy states $D_s\bar{D}_s/B_s\bar{B}_s$  with spin parity $J^P=0^+$, $D_s\bar{D}^*_s/B_s\bar{B}^*_s$ with $1^{+}$, and $D^*_s\bar{D}^*_s/B^*_s\bar{B}^*_s$ with $0^+$, $1^+$, and $2^+$.  In the doubly heavy sector, the bound states can be found from the interactions $(D^*D_s+DD^*_s)/(B^*B_s+BB^*_s)$ with $1^+$, $D_s\bar{D}_s^*/B_s\bar{B}_s^*$ with $1^+$, $D^*D^*_s/B^*B^*_s$ with $1^+$ and $2^+$, and $D^*_sD^*_s/B^*_sB^*_s$ with $1^+$ and $2^+$. Some other interactions are also found attractive, but may be not strong enough to produce a bound state. The results in this work are helpful for understanding the $Z_{cs}(3985)$, and future experimental search for the new molecular states. 
\end{abstract}

\section{Introduction}

After the observation of $X(3872)$ at Belle, more and more exotic hadrons are observed in experiment~\cite{Zyla:2020zbs}. How to understand the origin and internal structure of such states becomes an important central issue of hadron physics. With the accumulation of the experimental information, the molecular state becomes one of the most important pictures to explain these exotic states. For example, the hidden charm pentaquarks observed recently at LHCb are just a little below corresponding thresholds and exhibit an excellent spectrum of S-wave molecular states ~\cite{Aaij:2019vzc,Aaij:2015tga,Chen:2015loa,Roca:2015dva,Karliner:2015ina,He:2015cea,Liu:2019tjn,He:2019ify}. 

\begin{figure}[h!]
\includegraphics[bb=52 94 410 310,clip,scale=0.7]{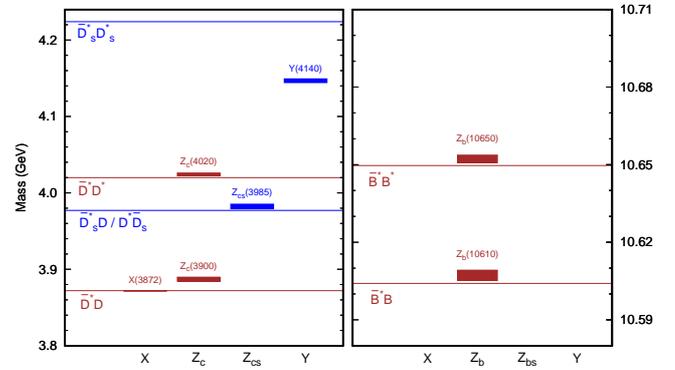}
\caption{The experimental masses of $X(3872)$, $Z_c(3900)$, $Z_c(4020)$, $Z_b(10610)$, $Z_c(10650)$, $Z_{cs}(3985)$, and $Y(4140)$ with relevant thresholds. The values are from the Review of Particle Physics~\cite{Zyla:2020zbs}.}
\label{XYZ}
\end{figure}

In the tetraquark sector, the first  exotic state $X(3872)$ was  soon explained as a molecular state due to its closeness to the $D\bar{D}^*$ threshold~\cite{Choi:2003ue,Tornqvist:2004qy,Swanson:2003tb}.  In the past decade, more and more structures were observed near the $D^{(*)}\bar{D}^{(*)}/B^{(*)}\bar{B}^{(*)}$ thresholds, such as the $Z_b(10610)$, $Z_b(10650)$, $Z_c(3900)$, and $Z_c(4020)$~\cite{Belle:2011aa,Ablikim:2013mio,Ablikim:2013wzq} (see Fig.~\ref{XYZ}). 
Recently, BESIII Collaboration announced observation of a new exotic candidate ${Z_{cs}}(3985)^-$ in $e^+$$e^-$ $\rightarrow$ ${K^+}{(D_s^-D^{*0} + D_s^{*-}D^0})$ near the ${D_{s}}^{-}{D}^{*0} / {D_{s}}^{*-}{D}^{0}$ threshold. It has a mass of $3982.5^{+1.8}_{-2.6}\pm 2.1$ MeV and a width of $12.8^{+5.3}_{-4.4}\pm 3.0$ MeV~\cite{Ablikim:2020hsk}. The LHCb also reported a structure $Z_{cs}(4000)$ at $4003\pm 6^{+4}_{-14}$~MeV, which is very close to the $Z_{cs}(3985)$,  but has a much broader width of $131\pm15\pm26$~MeV~\cite{Aaij:2021ivw}.  In 2009, the $Y(4140)$ state which is below the $D^*_s\bar{D}^*_s$  threshold was reported by the CDF Collaboration in the exclusive $B\to KJ/\psi\phi$ decay\cite{Aaltonen:2009tz,Aaltonen:2011at}, and confirmed at D0 and CMS in 2013~\cite{Chatrchyan:2013dma,Abazov:2013xda}.

Since the observation of these states, much theoretical efforts have been made to  understand their origin and internal structures.  Besides interpretation of the $X(3872)$ as a $D\bar{D}^*$  molecular state with quantum numbers $I^G(J^{PC})=0^-(1^{++})$, two $Z_b$ states are also widely taken as the candidates of the $B\bar{B}^*$ and $B^*\bar{B}^*$ molecular states~\cite{Sun:2011uh,Cleven:2011gp}. The $Z_c(3900)$ and $Z_c(4020)$ are good partners of the $Z_b$ states, and also be interpreted as molecular states 
of systems $D\bar{D}^*$ and $D^*\bar{D}^*$~\cite{Guo:2013sya,Wang:2013cya,Aceti:2014uea,He:2015mja,He:2013nwa}. However, As shown in Fig.~\ref{XYZ}, though these states are very close to the thresholds, the two $Z_b$ and two $Z_c$ states, especially the $Z_c(3900)$, are actually beyond the thresholds, which seems to conflict with the definition of the  molecular state as a loosely bound state. Moreover, many lattice calculations disfavor existence of a $D\bar{D}^*$ bound state~\cite{Prelovsek:2013xba,Prelovsek:2014swa,Ikeda:2016zwx,Chen:2014afa}. 
Therefore, many authors suggested that the $Z_c(3900)$ is a virtual state instead of a bound state from the $D\bar{D}^*$ interaction~\cite{Albaladejo:2015lob,He:2017lhy}. Since the virtual state locates on the second  Reimann sheet, its effect on the invariant mass spectrum should appear beyond the threshold.  As discussed in Ref.~\cite{Dong:2020hxe}, the attraction between two hadrons will exhibits itself as a structure even if it is not bound. Hence, the molecular state  interpretation of these states, especially if we categorize virtual state as a molecular state, is still the most reasonable and widely accepted~\cite{Dong:2021juy}. 

The observation of the hidden charm state with strangeness ${Z_{cs}}(3985)^-$ provides more confidence of the molecular interpretation of these charmonium-like states. Since the mass of the newly observed state is nearly 100 MeV larger than ${Z_{c}}(3900)$, which is the mass difference between ${D_{s}}^{*}$ and  ${D}^{*}$ meson, the state is recognized as a molecular state of ${D_{s}}^{-}{D}^{*0} + {D_{s}}^{*-}{D}^{0}$ in the literature~\cite{Sun:2020hjw,Meng:2020ihj,Wang:2020htx,Meng:2020cbk,Xu:2020evn,Wang:2020rcx,Ortega:2021enc,Chen:2021erj,Wang:2020dgr}. In Ref.~\cite{Meng:2020ihj}, the mass and width were well reproduced with an assignment of ${Z_{cs}}(3985)$  as a $U/V$-spin partner of ${Z_c}(3900)$. The states ${Z}^*_{cs}$, $Z_{bs}$ and ${Z}^*_{bs}$ were also predicted as the $U/V$-spin partner states of the charged ${Z_c}(4020)$, ${Z_b}(10610)$ and ${Z_b}(10650)$, respectively. Similar conclusion can be found in Ref~\cite{Wang:2020htx}, and the study was also extended to the $D^{(*)}_s\bar{D}^{(*)}_s$ systems~\cite{Meng:2020cbk}. In Ref.~\cite{Sun:2020hjw}, within a ${D_{s}}^{-}{D}^{*0}$ molecular state picture, authors found a pole located at $3982.34 -i0.53$~MeV on the third Reimann sheet of the complex energy plane by solving the Bethe-Salpeter equation with the on-shell factorization. The structure of $Y(4140)$ was also explained as a ${D}^{*}_{s}{\bar{D}}^{*}_{s}$ molecular state~\cite{Liu:2009ei,Albuquerque:2009ak,Wang:2014gwa,Zhang:2009st,Chen:2015fdn,Mahajan:2009pj}. In Ref.~\cite{Liu:2009ei}, the auothrs proposed that the $Y(4140)$ can be taken as a partner with hidden-strangeness of the $Y(3930)$. In Refs.~\cite{Albuquerque:2009ak,Zhang:2009st,Wang:2014gwa}, the scalar ${D}^{(*)}_{s}{\bar{D}}^{(*)}_{s}$ molecular state was studied with the QCD sum rules, and their result favors the assignment of the $Y (4140)$ as a molecular state. Chen $et~al$.  suggested that the $Y (4140)$ state should be considered as a mixed state of three pure molecule states ${D}^{*0}{\bar{D}}^{*0}$, ${D}^{*+}{{D}}^{*-}$ and ${D}^{*+}_{s}{{D}}^{*-}_{s}$~\cite{Chen:2015fdn}.  However, most molecular state interpretations suggested spin parity $0^+$, which is inconsistent with the quantum numbers determined at LHCb, $J^{PC}=1^{++}$~\cite{LHCb:2016axx}. The LHCb result disfavors the assignment of the $Y(4140)$ as an S-wave  ${D}^{(*)}_{s}{\bar{D}}^{(*)}_{s}$ molecular state. Anyway, the previous studies strongly suggests existence of  ${D}^{(*)}_{s}{\bar{D}}^{(*)}_{s}$ molecular states.   


In our previous studies, the $Z_c$ and $Z_b$ states were studied within the one-boson-exchange model by solving the Schr\"odinger equation or the quasipotential Bethe-Salpeter equation (qBSE)~\cite{Sun:2011uh,He:2014nya,He:2015mja}. If we extend such study to the $Z_{cs}$ state in the molecular state picture, a problem will appear. Different from the $D^{(*)}\bar{D}^{(*)}/B^{(*)}\bar{B}^{(*)}$ interaction,  no light meson can be exchanged between a charmed meson and a charmed strange meson. 
If we do not introduce the heavy meson exchange, no bound states can be produced from corresponding interactions. In Ref.~\cite{He:2015mja}, the $J/\psi$ exchange was included in the study of  $D^{(*)}\bar{D}^{(*)}$ system, and found essential to reproduce the $Z_c(3900)$ as in other approach~\cite{Aceti:2014uea}. And a further study of the systems $D^{(*)}\bar{B}^{(*)}/B^{(*)}\bar{D}^{(*)}$ and $D^{(*)}{B}^{(*)}/B^{(*)}{D}^{(*)}$ suggest that the $J/\psi$ or $\Upsilon$ exchange is really important to reproduce the experimentally observed states ${Z_{c}}(3900)$,  ${Z_{c}}(4020)$,  ${Z_{b}}(10610)$, and ${Z_{b}}(10650)$ ~\cite{Ding:2020dio}. 
The new experimental observation of ${Z_{cs}}(3985)$ inspires us to explore the possible molecular states from interactions $D^{(*)}{\bar{D}}^{(*)}_{s}$ / $B^{(*)}{\bar{B}}^{(*)}_{s}$,  ${D}^{(*)}_{s}{\bar{D}}^{(*)}_{s}$ / ${{B}}^{(*)}_{s}{\bar{B}}^{(*)}_{s}$, ${D}^{(*)}{D_{s}}^{(*)}$/${B}^{(*)}{B_{s}}^{(*)}$ and ${D_{s}}^{(*)}{D_{s}}^{(*)}$/${B_{s}}^{(*)}{B_{s}}^{(*)}$ with the $J/\psi$ or $\Upsilon$ exchange included. In the calculation, the potential of the interactions will be constructed with the meson exchange and inserted into the qBSE to obtain the scattering amplitudes.  The molecular states are searched for as the poles of the scattering amplitudes.

This article is organized as follows. After introduction, we present the details of theoretical frame in Sect.~\ref{Sec: Formalism}, which includes flavor wave functions, effective Lagrangians, construction of potential and a brief introduction of the qBSE approach. The numerical results for the states produced from the interaction considered will be given in Sect.~\ref{Sec: Results}. Finally, article ends with summary and discussion in Sect.~\ref{Sec: Summary}.

\section{Theoretical frame} \label{Sec: Formalism}

First, we should construct the  flavor wave functions. Since the systems considered in the current work contain one $u/d$ quark  at most, the isospins of states are fixed. We first present the wave functions for systems $D{\bar{D}}^*_{s}$ and $D^*{\bar{D}}^*_{s}$.  Since the thresholds of these two systems are very close, under the SU(3)$_F$ symmetry, the flavor wave functions for charged and neutral states can be expressed as ~\cite{Meng:2020ihj},
\begin{align}
|X_{D^*{\bar{D}}_{s}+D{\bar{D}}^*_{s}}^-\rangle&=\frac{1}{\sqrt{2}}\big(|{D}^{*0}D_{s}^{-}\rangle+\eta|{D}^{0}D_{s}^{*-}\rangle\big),\nonumber\\
|X_{D^*{\bar{D}}_{s}+D{\bar{D}}^*_{s}}^0\rangle&=\frac{1}{\sqrt{2}}\big(|{D}^{*+}D_{s}^{-}\rangle+\eta|{D}^{+}D_{s}^{*-}\rangle\big), \label{Eq: wf1}
\end{align}
where $\eta=\pm$  corresponds to the $Z_{cs}$ state with $G_{U/V}=\pm$.  To make the calculation practicable, the masses of $D_s$ and $D$, as well as $D^*_s$ and $D^*$ mesons,  will be chosen as their average value.  To check the effect of such treatment, we will also consider the direct wave functions without mixing as 
\begin{align}
|X_{{{D}}^*\bar{D}_s}^-\rangle&=|{D}^{*0}D_{s}^{-}\rangle,\quad
|X_{D{\bar{D}}^*_{s}}^-\rangle=|{D}^{0}D_{s}^{*-}\rangle,\nonumber\\
|X_{{{D}}^*\bar{D}_s}^0\rangle&=|{D}^{*+}D_{s}^{-}\rangle,\quad
|X_{D{\bar{D}}^*_{s}}^0\rangle=|{D}^{+}D_{s}^{*-}\rangle.
 \label{Eq: wf2}
\end{align}

For the ${D_{s}}\bar{D}_{s}^{*}$ states, the flavor function with a definite charge parity $C$ is constructed as
\begin{align}
|X_{{D_{s}}{\bar{D}}^*_{s}}^0\rangle&=\frac{1}{\sqrt{2}}\big(|D_{s}^{-}D_{s}^{*+}\rangle-C|D_{s}^{*-}D_{s}^{+}\rangle\big).
 \label{Eq: wf3}
\end{align}
Different from the SU(3)$_F$ symmetry, the $C$ parity is well defined because the masses of $D_s^+$ and $D_s^-$, as well as the masses of $D^{*+}_s$ and $D^{*-}_s$ mesons,  are almost the same. 

The flavor wave functions for hidden charm systems $D^*{\bar{D}}^*_{s}$ and ${D_{s}}{\bar{D}}_{s}$ can be written easily as 
\begin{align}
|X_{D^*{\bar{D}}^*_{s}}^-\rangle&=|{D}^{*0}D_{s}^{*-}\rangle, \quad |X_{D_{s}\bar{D}_{s}}^0\rangle=|D_{s}^{-}D_{s}^{+}\rangle,\nonumber\\
|X_{D^*{\bar{D}}^*_{s}}^0\rangle&=|{D}^{*+}D_{s}^{*-}\rangle. 
 \label{Eq: wf4}
\end{align}

For  doubly charm system with strangeness $D^*{D}_{s}/DD^*_s$, the closeness of the total masses of two channels also lead to strong mixing. Here, we construct the wave functions analogous to the $D^*{\bar{D}}_{s}/D\bar{D}^*_s$ as 
\begin{align}
|X_{D^*{D}_{s}+D{D}^*_{s}}^{+}\rangle&=\frac{1}{\sqrt{2}}\big(|{D}^{*0}D_{s}^{+}\rangle+\eta|{D}^{0}D_{s}^{*+}\rangle\big),\nonumber\\
|X_{D^*{D}_{s}+D{D}^*_{s}}^{++}\rangle&=\frac{1}{\sqrt{2}}\big(|{D}^{*+}D_{s}^{+}\rangle+\eta|{D}^{+}D_{s}^{*+}\rangle\big).
 \label{Eq: wf5}
\end{align}
Though this construction is not the standard U/V spin, we still  define a $G'=\pm$ which corresponds $\eta=\pm$. Direct wave function without mixing will be also considered as hidden charmed system  $D^*{\bar{D}}_{s}/D\bar{D}^*_s$.

The wave functions for $D^{*}{D_{s}}^{*}$ and $D{D_s}$ states can be constructed as,
\begin{align}
|X_{D^{*}D_{s}^{*}}^{+}\rangle&=|D^{*0}D_{s}^{*+}\rangle, \quad 
 |X_{D{D_{s}}}^{+}\rangle=|D^{0}D_{s}^{+}\rangle,\nonumber\\
 |X_{D^*D_s^*}^{++}\rangle&=|D^{*+}D_{s}^{*+}\rangle, \quad 
 |X_{D{D_{s}}}^{++}\rangle=|D^{+}D_{s}^{+}\rangle. \label{Eq: wf6}
\end{align}
For the doubly charm system with double strangeness $D_s^{(*)}D_s^{(*)}$, the wave function has a form of
\begin{align}
 |X_{D_s^{(*)}D_s^{(*)}}^{++}\rangle&=|D_{s}^{(*)+}D_{s}^{(*)+}\rangle. \label{Eq: wf7}
\end{align}
The wave functions for the hidden bottom and doubly bottom systems can be obtained analogously. 

To achieve the potential kernel for the interactions, the meson exchange model will be adopted.  The Lagrangians under the chiral symmetry and heavy quark limit will be introduced to depict the vertices of the meson exchanges. The Lagrangians for heavy mesons interacting with light mesons read~\cite{Cheng:1992xi,Yan:1992gz,Wise:1992hn,Burdman:1992gh,Casalbuoni:1996pg},
\begin{align}
\mathcal{L}_{\mathcal{P}^*\mathcal{P}^*\mathbb{P}} &=
-i\frac{2g}{f_\pi}\varepsilon_{\alpha\mu\nu\lambda}
v^\alpha\mathcal{P}^{*\mu}_{b}{\mathcal{P}}^{*\lambda\dag}_{a}
\partial^\nu{}\mathbb{P}_{ba}\nonumber\\
&+i \frac{2g}{f_\pi}\varepsilon_{\alpha\mu\nu\lambda}
v^\alpha\widetilde{\mathcal{P}}^{*\mu\dag}_{a}\widetilde{\mathcal{P}}^{*\lambda}_{b}
\partial^\nu{}\mathbb{P}_{ab},\nonumber\\
\mathcal{L}_{\mathcal{P}^*\mathcal{P}\mathbb{P}} &=-
\frac{2g}{f_\pi}(\mathcal{P}^{}_b\mathcal{P}^{*\dag}_{a\lambda}+
\mathcal{P}^{*}_{b\lambda}\mathcal{P}^{\dag}_{a})\partial^\lambda{}
\mathbb{P}_{ba}\nonumber\\
&+\frac{2g}{f_\pi}(\widetilde{\mathcal{P}}^{*\dag}_{a\lambda}\widetilde{\mathcal{P}}_b+
\widetilde{\mathcal{P}}^{\dag}_{a}\widetilde{\mathcal{P}}^{*}_{b\lambda})\partial^\lambda{}\mathbb{P}_{ab},
\nonumber\\
  \mathcal{L}_{\mathcal{PP}\mathbb{V}}
  &= -\sqrt{2}\beta{}g_V\mathcal{P}^{}_b\mathcal{P}_a^{\dag}
  v\cdot\mathbb{V}_{ba}
 +\sqrt{2}\beta{}g_V\widetilde{\mathcal{P}}^{\dag}_a
  \widetilde{\mathcal{P}}^{}_b
  v\cdot\mathbb{V}_{ab},\nonumber\\
  \mathcal{L}_{\mathcal{P}^*\mathcal{P}\mathbb{V}}
  &=- 2\sqrt{2}\lambda{}g_V v^\lambda\varepsilon_{\lambda\mu\alpha\beta}
  (\mathcal{P}^{}_b\mathcal{P}^{*\mu\dag}_a +
  \mathcal{P}_b^{*\mu}\mathcal{P}^{\dag}_a)
  (\partial^\alpha{}\mathbb{V}^\beta)_{ba}\nonumber\\
&-  2\sqrt{2}\lambda{}g_V
v^\lambda\varepsilon_{\lambda\mu\alpha\beta}
(\widetilde{\mathcal{P}}^{*\mu\dag}_a\widetilde{\mathcal{P}}^{}_b
+
\widetilde{\mathcal{P}}^{\dag}_a\widetilde{\mathcal{P}}_b^{*\mu})
  (\partial^\alpha{}\mathbb{V}^\beta)_{ab},\nonumber\\
  \mathcal{L}_{\mathcal{P}^*\mathcal{P}^*\mathbb{V}}
  &= \sqrt{2}\beta{}g_V \mathcal{P}_b^{*}\cdot\mathcal{P}^{*\dag}_a
  v\cdot\mathbb{V}_{ba}\nonumber\\
  &-i2\sqrt{2}\lambda{}g_V\mathcal{P}^{*\mu}_b\mathcal{P}^{*\nu\dag}_a
  (\partial_\mu{}
  \mathbb{V}_\nu - \partial_\nu{}\mathbb{V}_\mu)_{ba}\nonumber\\
  &-\sqrt{2}\beta g_V
  \widetilde{\mathcal{P}}^{*\dag}_a\widetilde{\mathcal{P}}_b^{*}
  v\cdot\mathbb{V}_{ab}\nonumber\\
  &-i2\sqrt{2}\lambda{}g_V\widetilde{\mathcal{P}}^{*\mu\dag}_a\widetilde{\mathcal{P}}^{*\nu}_b(\partial_\mu{}
  \mathbb{V}_\nu - \partial_\nu{}\mathbb{V}_\mu)_{ab},
\nonumber\\
  \mathcal{L}_{\mathcal{PP}\mathbb{V}}
  &= -\sqrt{2}\beta{}g_V\mathcal{P}^{}_b\mathcal{P}_a^{\dag}
  v\cdot\mathbb{V}_{ba}
 +\sqrt{2}\beta{}g_V\widetilde{\mathcal{P}}^{\dag}_a
  \widetilde{\mathcal{P}}^{}_b
  v\cdot\mathbb{V}_{ab},\nonumber\\
  \mathcal{L}_{\mathcal{P}^*\mathcal{P}\mathbb{V}}
  &=- 2\sqrt{2}\lambda{}g_V v^\lambda\varepsilon_{\lambda\mu\alpha\beta}
  (\mathcal{P}^{}_b\mathcal{P}^{*\mu\dag}_a +
  \mathcal{P}_b^{*\mu}\mathcal{P}^{\dag}_a)
  (\partial^\alpha{}\mathbb{V}^\beta)_{ba}\nonumber\\
&-  2\sqrt{2}\lambda{}g_V
v^\lambda\varepsilon_{\lambda\mu\alpha\beta}
(\widetilde{\mathcal{P}}^{*\mu\dag}_a\widetilde{\mathcal{P}}^{}_b
+
\widetilde{\mathcal{P}}^{\dag}_a\widetilde{\mathcal{P}}_b^{*\mu})
  (\partial^\alpha{}\mathbb{V}^\beta)_{ab},\nonumber\\
  \mathcal{L}_{\mathcal{P}^*\mathcal{P}^*\mathbb{V}}
  &= \sqrt{2}\beta{}g_V \mathcal{P}_b^{*}\cdot\mathcal{P}^{*\dag}_a
  v\cdot\mathbb{V}_{ba}\nonumber\\
  &-i2\sqrt{2}\lambda{}g_V\mathcal{P}^{*\mu}_b\mathcal{P}^{*\nu\dag}_a
  (\partial_\mu{}
  \mathbb{V}_\nu - \partial_\nu{}\mathbb{V}_\mu)_{ba}\nonumber\\
  &-\sqrt{2}\beta g_V
  \widetilde{\mathcal{P}}^{*\dag}_a\widetilde{\mathcal{P}}_b^{*}
  v\cdot\mathbb{V}_{ab}\nonumber\\
  &-i2\sqrt{2}\lambda{}g_V\widetilde{\mathcal{P}}^{*\mu\dag}_a\widetilde{\mathcal{P}}^{*\nu}_b(\partial_\mu{}
  \mathbb{V}_\nu - \partial_\nu{}\mathbb{V}_\mu)_{ab},
\nonumber\\
  \mathcal{L}_{\mathcal{PP}\sigma}
  &= -2g_s\mathcal{P}^{}_b\mathcal{P}^{\dag}_b\sigma
 -2g_s\widetilde{\mathcal{P}}^{}_b\widetilde{\mathcal{P}}^{\dag}_b\sigma,\nonumber\\
  \mathcal{L}_{\mathcal{P}^*\mathcal{P}^*\sigma}
  &= 2g_s\mathcal{P}^{*}_b\cdot{}\mathcal{P}^{*\dag}_b\sigma
 +2g_s\widetilde{\mathcal{P}}^{*}_b\cdot{}\widetilde{\mathcal{P}}^{*\dag}_b\sigma, \label{Eq: L}
\end{align} 
where  the velocity $v$ should be replaced by $i\overleftrightarrow{\partial}/2\sqrt{m_im_f}$ with the $m_{i,f}$ being the mass of the initial or final heavy meson. 
${\mathcal{P}}^{(*)T} =(D^{(*)0},D^{(*)+},D_s^{(*)+})$ or
$(B^{(*)-},\bar{B}^{(*)0},\bar{B}_s^{(*)0})$, and 
 satisfy the normalization relations $\langle
0|{\mathcal{P}}|{Q}\bar{q}(0^-)\rangle
=\sqrt{M_\mathcal{P}}$ and $\langle
0|{{\mathcal{P}}}^*_\mu|{Q}\bar{q}(1^-)\rangle=
\epsilon_\mu\sqrt{M_{\mathcal{P}^*}}$. 
The $\mathbb
P$ and $\mathbb V$ denote the pseudoscalar and vector matrices
\begin{equation}
    {\mathbb P}=\left(\begin{array}{ccc}
        \frac{\sqrt{3}\pi^0+\eta}{\sqrt{6}}&\pi^+&K^+\\
        \pi^-&\frac{-\sqrt{3}\pi^0+\eta}{\sqrt{6}}&K^0\\
        K^-&\bar{K}^0&-\frac{2\eta}{\sqrt{6}}
\end{array}\right),\ \ 
\mathbb{V}=\left(\begin{array}{ccc}
\frac{\rho^0+\omega}{\sqrt{2}}&\rho^{+}&K^{*+}\\
\rho^{-}&\frac{-\rho^{0}+\omega}{\sqrt{2}}&K^{*0}\\
K^{*-}&\bar{K}^{*0}&\phi
\end{array}\right).\label{MPV}
\end{equation}
The parameters involved here were determined in the literature as $g=0.59$, $\beta=0.9$, $\lambda=0.56$ GeV$^{-1}$, and $g_s=0.76$ with $f_\pi=132$ MeV~\cite{Falk:1992cx,Isola:2003fh,Liu:2008tn,Chen:2019asm}.

In Refs.~\cite{Aceti:2014uea,He:2015mja}, contribution from the $J/\psi$ exchange is found important in the $D\bar{D}^*$ interaction to produce the $Z_c(3900)$ observed at BESIII.  In the current work, we also consider such exchange with the couplings of heavy-light charmed mesons to $J/\psi$, which are written with the help of  heavy quark effective theory as \cite{Casalbuoni:1996pg,Oh:2000qr},
\begin{eqnarray}
	{\cal L}_{D^*_{(s)}\bar{D}^*_{(s)}J/\psi}&=&-ig_{D^*_{(s)}D^*_{(s)}\psi}\big[\psi \cdot \bar{D}^*\overleftrightarrow{\partial}\cdot D^*\nonumber\\
&-&
\psi^\mu \bar D^* \cdot\overleftrightarrow{\partial}^\mu {D}^* +
\psi^\mu \bar{D}^*\cdot\overleftrightarrow{\partial} D^{*\mu} ) \big], \nonumber \\
{\cal L}_{D_{(s)}^*\bar{D}_{(s)}J/\psi}&=&
g_{D^*_{(s)}D_{(s)}\psi} \,  \, \epsilon_{\beta \mu \alpha \tau}
\partial^\beta \psi^\mu (\bar{D}
\overleftrightarrow{\partial}^\tau D^{* \alpha}+\bar{D}^{* \alpha}
\overleftrightarrow{\partial}^\tau D) \label{matrix3}, \nonumber \\
{\cal L}_{D_{(s)} \bar{D}_{(s)}J/\psi} &=&
ig_{D_{(s)}D_{(s)}\psi} \psi \cdot
\bar{D}\overleftrightarrow{\partial}D,
\end{eqnarray}
where the couplings
are related to a single parameter $g_2$ as
\begin{eqnarray}
\frac{g_{D^*D^*\psi} }{m_{D^*}}= \frac{g_{D_{(s)}D_{(s)}\psi}}{m_D}= g_{D^*_{(s)}D_{(s)}\psi}= 2 g_2 \sqrt{m_\psi },\label{Eq: para}
\end{eqnarray}
with $g_2={\sqrt{m_\psi}}/({2m_Df_\psi})$ and $f_\psi=405$ MeV.  For the bottom mesons,  analogous Lagrangians can be obtained under the heavy quark symmetry for  $\Upsilon$ exchange. The parameters can be obtained by similar relation in Eq.~(\ref{Eq: para}) by replacing the mass by these of bottom mesons with $f_\Upsilon=715$ MeV~\cite{Li:2012as}. 

With the help of  standard Feynman rule,  the vertices can be easily obtained from the above Lagrangians. The potential  interaction can be constructed as~\cite{He:2019ify},
\begin{equation}%
{\cal V}_{\mathbb{P},\sigma}=I^{d,c}_i\Gamma_1\Gamma_2 P_{\mathbb{P},\sigma}f(q^2),\ \ 
{\cal V}_{\mathbb{V}}=I^{d,c}_i\Gamma_{1\mu}\Gamma_{2\nu}  P^{\mu\nu}_{\mathbb{V}}f(q^2),\label{V}
\end{equation}
where the propagators are defined as usual as
\begin{equation}%
P_{\mathbb{P},\sigma}= \frac{i}{q^2-m_{\mathbb{P},\sigma}^2},\ \
P^{\mu\nu}_\mathbb{V}=i\frac{-g^{\mu\nu}+q^\mu q^\nu/m^2_{\mathbb{V}}}{q^2-m_\mathbb{V}^2}.
\end{equation}
We introduce a form factor $f(q^2)=\Lambda_e^2/(q^2-\Lambda_e^2)$  to reflect the off-shell effect of exchanged meson 
with $m_e$ being the $m_{\mathbb{P},\mathbb{V},\sigma}$ and $q$ being the momentum of the exchanged  meson. The current form factor can avoid overestimation of the contribution of $J/\psi$ exchange.

In our approach, we collect the coefficients for the interaction of  states as  flavor factors $I^{d,c}_i$.  As discussed in Refs.~\cite{He:2014nya,Ding:2020dio}, for the hidden heavy state, the cross diagram  appears in the $[D{{\bar{D}}^{*}_{s}}]$/$[B{{\bar{B}}^{*}_{s}}]$  and ${{D}}_{s}{\bar{D}}^*_{s}$/${{B}}_{s}{\bar{B}}^*_{s}$ cases due to the coupling between the two parts as shown in Eq.~(\ref{Eq: wf1}) and Eq.~(\ref{Eq: wf3}).  For the states  $D{\bar{D}_{s}}/B{\bar{B}_{s}}$, $D^{*}{{\bar{D}}^{*}_{s}}/B^{*}{{\bar{B}}^{*}_{s}}$, $D_{s}{\bar{D}_{s}}/B_{s}{\bar{B}_{s}}$ and $D_{s}^{*}{{\bar{D}}^{*}_{s}} / B_{s}^{*}{{\bar{B}}^{*}_{s}}$, there is no cross diagram.  For the doubly heavy states,  there is no cross diagram from the coupling in  the ${D}^{(*)}{D_{s}}^{(*)}$/${B}^{(*)}{B_{s}}^{(*)}$ and the ${D_{s}}^{(*)}{D_{s}}^{(*)}$/${B_{s}}^{(*)}{B_{s}}^{(*)}$ cases. However, in such cases, the $u$ channel is allowed, which   provides cross diagram.  In Table~\ref{Tab: flavor factor},  flavor factors $I^d_i$ and $I^c_i$ of certain meson exchange $i$ of certain interaction are listed for direct and cross diagrams, respectively.
\renewcommand\tabcolsep{0.05cm}
\renewcommand{\arraystretch}{1.5}
\begin{table}[h!]
\begin{center}
\caption{The isospin factors $I_i^d$ and $I_i^c$ for direct and cross diagrams and different exchange mesons. Here we use notations ${\mathcal{P}}^{(*)} =D^{(*)}$ or
$B^{(*)}$,  ${\mathcal{P}}^{(*)}_s =D^{(*)}_s$ or
$B^{(*)}_s$, $[\mathcal{P}{\bar{\mathcal P}}_{s}^{*}]=\mathcal{P}^*{\bar{\mathcal{P}}}_{s}+\mathcal{P}{\bar{\mathcal{P}}}^*_{s}$, $[\mathcal{P}{{\mathcal P}}_{s}^{*}]=\mathcal{P}^*{{\mathcal{P}}}_{s}+\mathcal{P}{{\mathcal{P}}}^*_{s}$.}
\label{Tab: flavor factor}
\begin{tabular}{ccccccccccccc}\bottomrule[2pt]\hline
Channel & \multicolumn{5}{c}{$I_i^d$}& \multicolumn{5}{c}{ $I_i^c$}\\\cmidrule(lr){1-1}\cmidrule(lr){2-6}\cmidrule(lr){7-11}
&$K$&$\eta$  &$K^{*}$ &$\phi$&$J/\psi$ &$K$&$\eta$  &$K^{*}$ &$\phi$&$J/\psi$ \\
$[\mathcal{P}{\bar{\mathcal{P}}}_{s}^{*}]$&$\tdots$&$\tdots$&$\tdots$ &$\tdots$ &$1$
&$\tdots$&$\tdots$&$\tdots$ &$\tdots$ &$\eta$\\
$\mathcal{P}_{s}{\bar{\mathcal{P}}}_{s}^{*}$&$\tdots$&$\tdots$&$\tdots$&1&$1$
&$\tdots$&$-\frac{2}{3}C$&$\tdots$ &$-C$ &$-C$\\
$\mathcal{P}_{s}{\bar{\mathcal{P}}}_{s} \rightarrow [\mathcal{P}_{s}{\bar{\mathcal{P}}}_{s}^{*}]$&$\tdots$&$\tdots$&$\tdots$&$\frac{1}{\sqrt{2}}$&$ \frac{1}{\sqrt{2}}$
&$\tdots$&$\tdots$&$\tdots$ &$-\frac{1}{\sqrt{2}}C$ &$ -\frac{1}{\sqrt{2}}C$\\
$[\mathcal{P}_{s}{\bar{\mathcal{P}}}_{s}^{*}]\rightarrow \mathcal{P}_{s}^{*}{\bar{\mathcal{P}}}_{s}^{*}$&$\tdots$&$\frac{\sqrt{2}}{3}$&$\tdots$ &$\frac{1}{\sqrt{2}}$ & $\frac{1}{\sqrt{2}}$
&$\tdots$&$-\frac{\sqrt{2}}{3}C$&$\tdots$ &$-\frac{1}{\sqrt{2}}C$ &$ -\frac{1}{\sqrt{2}}C$\\\cmidrule(lr){1-1}\cmidrule(lr){2-6}\cmidrule(lr){7-11}
$\mathcal{P}^{(*)}\bar{\mathcal{P}}_{s}^{(*)} \rightarrow \mathcal{P}^{(*)} \bar{\mathcal{P}}_{s}^{(*)}$&$\tdots$&$\tdots$&$\tdots$&$\tdots$&$1$
&$\tdots$&$\tdots$&$\tdots$&$\tdots$  &$\tdots$\\
$\mathcal{P}_{s}^{(*)}{\bar{\mathcal{P}}}_{s}^{(*)}\rightarrow \mathcal{P}_{s}^{(*)}{\bar{\mathcal{P}}}_{s}^{(*)}$&$\tdots$&$[\frac{2}{3}]$&$\tdots$&$1$&$1$
&$\tdots$&$\tdots$&$\tdots$ &$\tdots$  &$\tdots$\\
\cmidrule(lr){1-1}\cmidrule(lr){2-6}\cmidrule(lr){7-11}
$[\mathcal{P} \mathcal{P}_{s}^{*}] \rightarrow [\mathcal{P}\mathcal{P}_{s}^{*}]$&$\tdots$&$\tdots$&$\eta$&$\tdots$&$1$
&$1$&$\tdots$&$1$&$\tdots$&$\eta$\\
\cmidrule(lr){1-1}\cmidrule(lr){2-6}\cmidrule(lr){7-11}
$\mathcal{P}^{(*)} \mathcal{P}_{s}^{(*)} \rightarrow \mathcal{P}^{(*)}\mathcal{P}_{s}^{(*)}$&$\tdots$&$\tdots$&$\tdots$&$\tdots$&$1$
&$[1]$&$\tdots$&$1$&$\tdots$&$\tdots$\\
$\mathcal{P}_{s}^{(*)} \mathcal{P}_{s}^{(*)} \rightarrow \mathcal{P}_{s} ^{(*)}\mathcal{P}_{s}^{(*)}$&$\tdots$&$[\frac{2}{3}]$&$\tdots$&$1$&$1 $
&$\tdots$&$[\frac{2}{3}]$&$\tdots$&$1$&$ 1$\\\hline
\toprule[2pt]
\end{tabular}
\end{center}
\end{table}

In the above, we construct the potential of the interactions considered in the current work. The scattering amplitude can be obtained with the qBSE~\cite{He:2014nya,He:2015mja,He:2017lhy,He:2015yva,He:2017aps}. After  the partial-wave decomposition,  the qBSE can be reduced to a 1-dimensional  equation with a spin-parity $J^P$ as~\cite{He:2015mja},
\begin{align}
i{\cal M}^{J^P}_{\lambda'\lambda}({\rm p}',{\rm p})
&=i{\cal V}^{J^P}_{\lambda',\lambda}({\rm p}',{\rm
p})+\sum_{\lambda''}\int\frac{{\rm
p}''^2d{\rm p}''}{(2\pi)^3}\nonumber\\
&\cdot
i{\cal V}^{J^P}_{\lambda'\lambda''}({\rm p}',{\rm p}'')
G_0({\rm p}'')i{\cal M}^{J^P}_{\lambda''\lambda}({\rm p}'',{\rm
p}),\quad\quad \label{Eq: BS_PWA}
\end{align}
where the ${\cal M}^{J^P}({\rm p}',{\rm p})$ is partial-wave scattering amplitude, and
the $G_0({\rm p}'')$ is reduced propagator with the spectator approximation in the center-of-mass frame as~\cite{He:2015mja},
\begin{eqnarray}
	G_0&=&\frac{\delta^+(p''^{~2}_h-m_h^{2})}{p''^{~2}_l-m_l^{2}}=\frac{\delta^+(p''^{0}_h-E_h)}{2E_h[(W-E_h)^2-E_l^{2}]}.
\end{eqnarray}
As required by the spectator approximation, the heavier particle  (remarked with $h$) satisfies $p''^0_h=E_{h}({\rm p}'')=\sqrt{
m_{h}^{~2}+\rm p''^2}$. The $p''^0_l$ for the lighter particle (remarked as $l$) is then $W-E_{h}({\rm p}'')$ with $W$ being the center-of-mass energy of the system. Here and hereafter, we define the value of the momentum  in center-of-mass frame as ${\rm p}=|{\bm p}|$.

The partial wave potential can be obtained from the potential in Eq.~(\ref{V}) as
\begin{align}
{\cal V}_{\lambda'\lambda}^{J^P}({\rm p}',{\rm p})
&=2\pi\int d\cos\theta
~[d^{J}_{\lambda\lambda'}(\theta)
{\cal V}_{\lambda'\lambda}({\bm p}',{\bm p})\nonumber\\
&+\eta d^{J}_{-\lambda\lambda'}(\theta)
{\cal V}_{\lambda'-\lambda}({\bm p}',{\bm p})],
\end{align}
where $\eta=PP_1P_2(-1)^{J-J_1-J_2}$ with $P$ and $J$ being parity and spin for system,  and two constituent heavy mesons. The initial and final relative momenta are chosen as ${\bm p}=(0,0,{\rm p})$  and ${\bm p}'=({\rm p}'\sin\theta,0,{\rm p}'\cos\theta)$. The $d^J_{\lambda\lambda'}(\theta)$ is the Wigner d-matrix.
An  exponential
regularization  was also introduced as  a form factor into the reduced propagator as $G_0({\rm p}'')\to G_0({\rm p}'')e^{-2(p''^2_l-m_l^2)^2/\Lambda_r^4}$~\cite{He:2015mja}.

\section{Numerical results}\label{Sec: Results}

With the preparation above, we can obtain the scattering amplitudes of the interaction of hidden heavy and doubly heavy systems, and the molecular states can be searched for as the poles in the complex energy plane. Because we are only interesting in the pole of the scattering amplitude,  we only need to find the  position where $|1-V(z)G(z)|=0$ with $z$ being the complex continuation of the center-of-mass energy of the system $W$ ~\cite{He:2015mja}.  In addition, we take two free parameters $\Lambda_e$ and $\Lambda_r$ as $\Lambda$ for simplification. 

\subsection{Numerical results with single-channel calculation}

In Tables~\ref{Tab: DD bound state}-\ref{Tab: DADA bound state}, we present the results for the  binding energy $E_B$ with single-channel calculation. Here,  the binding energy is defined as $E_B=M_{th}-z$, with the $M_{th}$  and $z$ being the threshold and  the position of the pole of the bound state. For single-channel calculation, the pole is at  real axis of the complex energy plane.   In the current work, we only consider the spin parities which can be produced in S-wave. Here we scan the values of cutoff $\Lambda$ in a range smaller than 5 GeV and present results with some selected  values of cutoff $\Lambda$ if there is a bound state produced from the corresponding interaction with a binding energy smaller than 40~MeV.

The results for the systems of two pseudoscalar mesons are listed in Table~\ref{Tab: DD bound state}.  
Only state with spin parity $0^+$ can be produced in S-wave for  two pseudoscalar mesons.  In the charmed sector, the bound state can be produced from the hidden charmed systems $D\bar{D}_s$ and $D_s\bar{D}_s$. However the cutoff required to  produce the former state is much larger than the later one. For the hidden bottom systems $B\bar{B}_s$ and $B_s\bar{B}_s$, the interactions are still attractive,  but a  cutoff larger than 5 GeV is required to produce a $B\bar{B}_s$ bound state. The attraction in the $B\bar{B}_s$ interaction is strong enough to form a bound state at a cutoff of about 1~GeV. No bound states is found in the doubly charmed and doubly bottom systems, $DD_s$, $D_sD_s$, $BB_s$, and $B_sB_s$. Hence, our results support the existence of  deeply bound states $D_s\bar{D}_s$ and $B_s\bar{B}$ with $0^+$, and the $D\bar{D}_s$ interaction are considerably attractive.

\renewcommand\tabcolsep{0.28cm}
\renewcommand{\arraystretch}{1.}
\begin{table}[hbtp!]
\begin{center}
\caption{The binding energies $E_B=M_{th}-z$ of the bound states from the interactions $D_{(s)}\bar{D}_s/B_{(s)}\bar{B}_s$ and $D_{(s)}\bar{D}_s/B_{(s)}\bar{B}_s$  with some selected values of cutoff $\Lambda$. The ``$\tdots$" means that no bound state is found in the considered range of the cutoff. The cutoff $\Lambda$ and binding energy $W$ are in the units of GeV and MeV.
\label{Tab: DD bound state}
\label{diagrams}}
\begin{tabular}{ccccccccccc}\bottomrule[2pt]\hline
$$ & \multicolumn{2}{c}{$D{\bar{D}}_{s}$}& \multicolumn{2}{c}{$D_{s}{\bar{D}}_{s}$}&\multicolumn{2}{c}{$DD_{s}$}&\multicolumn{2}{c}{$D_{s}D_{s}$} \\\cmidrule(lr){2-3}\cmidrule(lr){4-5}\cmidrule(lr){6-7}\cmidrule(lr){8-9}
$J^{P}$   &  $\Lambda$ & $E_B$ & $\Lambda$ &$E_B$ & $\Lambda$ & $E_B$&  $\Lambda$ & $E_B$ \\
$0^{+}$& $4.4$ & $1.1$& $1.6$ & $0.3$&$\tdots$ &$\tdots$ &$\tdots$ &$\tdots$ \\
             & $4.7$ & $8.8$ & $2.1$ & $8.4$& $\tdots$ &  $\tdots$ &$\tdots$&$\tdots$ \\
              & $4.9$ & $18.4$ & $2.4$ & $19.5$ & $\tdots$ &$\tdots$ &$\tdots$ &$\tdots$\\
              \hline
 $$ & \multicolumn{2}{c}{$B{\bar{B}}_{s}$}& \multicolumn{2}{c}{$B_{s}{\bar{B}}_{s}$}&\multicolumn{2}{c}{$BB_{s}$}&\multicolumn{2}{c}{$B_{s}B_{s}$} \\\cmidrule(lr){2-3}\cmidrule(lr){4-5}\cmidrule(lr){6-7}\cmidrule(lr){8-9}
$J^{P}$   &  $\Lambda$ & $E_B$ & $\Lambda$ & $E_B$ & $\Lambda$ & $E_B$ &  $\Lambda$ & $E_B$ \\
$0^{+}$&$\tdots$ & $\tdots$& $1.0$ & $1.4$&$\tdots$ &$\tdots$ &$\tdots$ &$\tdots$ \\
            & $\tdots$ & $\tdots$ & $1.5$ & $9.8$ & $\tdots$ &  $\tdots$ &$\tdots$&$\tdots$ \\
              & $\tdots$ & $\tdots$ & $1.8$ & $18.7$ & $\tdots$ &$\tdots$ &$\tdots$ &$\tdots$\\\hline
\toprule[2pt]
\end{tabular}
\end{center}

\end{table}

The results for the systems of a pseudoscalar meson and a vector meson are listed in Table~\ref{Tab: DDA bound state}.  
The spin parities for these systems are $1^+$ in S wave. Here, the bound states with different $G_{U/V}$ spins introduced in Eq.~(\ref{Eq: wf1}) are listed for the hidden charmed strange system $D^*{\bar{D}}_{s}+D\bar{D}^*_s$. The calculation supports the existence of a state with $G_{U/V}=+$ at a cutoff of about 3 GeV, which  is relevant to the state $Z_{cs}(3985)$ observed at BESIII recently. No bound state can be found with $G_{U/V}=-$ from the $D^*{\bar{D}}_{s}+D\bar{D}^*_s$ interaction. In the bottom sector, the $B^*{\bar{B}}_{s}+B\bar{B}^*_s$ interaction with  $G_{U/V}=+$ is still attractive while a  cutoff larger than 5~GeV is required to produce a bound state. There is still no bound state produced from the interaction with $G_{U/V}=-$. 
For the doubly charmed system, we still consider the system with $G'$, a $D^*{{D}}_{s}+DD_{s}^*$ state with $G'=-$ can be found at a cutoff of about 2~GeV, and a state with the same $G'$ spin is produced from the interaction $B^*{{B}}_{s}+BB_{s}^*$ at a smaller cutoff of about 0.8~GeV. 

\renewcommand\tabcolsep{0.235cm}
\renewcommand{\arraystretch}{1.}
\begin{table}[hbtp!]
\begin{center}
\caption{The binding energies of the bound states from the interactions $D_{(s)}\bar{D}^*_s/B_{(s)}\bar{B}^*_s$, and $D_{(s)}{D}^*_s/B_{(s)}{B}^*_s$  with some selected values of cutoff $\Lambda$. The $G_{U/V}$, $G'$, and $C$ are defined in Eqs.~(\ref{Eq: wf1}, \ref{Eq: wf3}, \ref{Eq: wf5}). Other notations are the same as Table~\ref{Tab: DD bound state}.
\label{Tab: DDA bound state}
\label{diagrams}}
\begin{tabular}{ccccccccccc}\bottomrule[2pt]\hline
$$& \multicolumn{4}{c}{$D^*{\bar{D}}_{s}+D\bar{D}^{*}_{s}$} &\multicolumn{4}{c}{$D^*{{D}}_{s}+DD_{s}^*$}	 \\\cmidrule(lr){2-5}\cmidrule(lr){6-9}
$$& \multicolumn{2}{c}{$G_{U/V}=+$}& \multicolumn{2}{c}{$G_{U/V}=-$} &\multicolumn{2}{c}{$G'=+$}&\multicolumn{2}{c}{$G'=-$}	 \\\cmidrule(lr){2-3}\cmidrule(lr){4-5}\cmidrule(lr){6-7}\cmidrule(lr){8-9}
$J^{P}$   &  $\Lambda$ & $E_B$ & $\Lambda$ & $E_B$ & $\Lambda$ & $E_B$ & $\Lambda$ & $E_B$\\
$1^{+}$& 3.1 & \ \ 1.4&$\tdots$&$\tdots$&$\tdots$&$\tdots$&$2.0$ &\ \ 0.4\\
             & 3.3 & 10.8&$\tdots$&$\tdots$&$\tdots$&$\tdots$& 2.3 & 10.6 \\
              & 3.4 & 19.1&$\tdots$&$\tdots$ &$\tdots$&$\tdots$& 2.4  & 17.7\\
             \hline
             $$& \multicolumn{4}{c}{$D_s{\bar{D}}^{*}_{s}$} &\multicolumn{2}{c}{$D_sD_{s}^*$}	 \\\cmidrule(lr){2-5}\cmidrule(lr){6-7}
$$& \multicolumn{2}{c}{$C=+$}& \multicolumn{2}{c}{$C=-$} &	\multicolumn{2}{c}{$--$}  \\\cmidrule(lr){2-3}\cmidrule(lr){4-5}\cmidrule(lr){6-7}
 $J^{P}$   &  $\Lambda$ & $E_B$ & $\Lambda$ & $E_B$ & $\Lambda$ & $E_B$ \\    
$1^{+}$& 1.4 & \ \ 0.8& 1.8&\ \ 1.0&$3.40$ &\ \ $0.1$ \\
             &$1.7$ &\ \  $9.7$ &2.3&\ \ 7.8&$3.42$&$10.5$\\
              &$1.8$ & $15.8$&2.6&16.8&$3.43$ &$18.1$ \\\hline
$$& \multicolumn{4}{c}{$B^*{\bar{B}}_{s}+B{\bar{B}}^{*}_{s}$} &\multicolumn{4}{c}{$B^*{{B}}_{s}+BB_{s}^*$}	 \\\cmidrule(lr){2-5}\cmidrule(lr){6-9}
$$& \multicolumn{2}{c}{$G_{U/V}=+$}& \multicolumn{2}{c}{$G_{U/V}=-$} &\multicolumn{2}{c}{$G'=+$}&\multicolumn{2}{c}{$G'=-$}	 \\\cmidrule(lr){2-3}\cmidrule(lr){4-5}\cmidrule(lr){6-7}\cmidrule(lr){8-9}
$J^{P}$   &  $\Lambda$ & $E_B$ & $\Lambda$ & $E_B$ & $\Lambda$ & $E_B$& $\Lambda$ & $E_B$ \\
$1^{+}$ &$\tdots$ &$ \tdots$&$\tdots$&$\tdots$&$\tdots$&$\tdots$&$0.8$ & 0.9\\
             & $\tdots$& $\tdots$&$\tdots$&$\tdots$&$\tdots$&$\tdots$& 1.2 & 10.0\\
            & $\tdots$ &$\tdots$&$\tdots$&$\tdots$ &$\tdots$&$\tdots$ & 1.4 & 19.6 \\
             \hline
              $$& \multicolumn{4}{c}{$B_s{\bar{B}}^{*}_{s}$} &\multicolumn{2}{c}{$B_sB_{s}^*$}	 \\\cmidrule(lr){2-5}\cmidrule(lr){6-7}
$$& \multicolumn{2}{c}{$C=+$}& \multicolumn{2}{c}{$C=-$} & \multicolumn{2}{c}{$--$} 	 \\\cmidrule(lr){2-3}\cmidrule(lr){4-5}\cmidrule(lr){6-7}
$J^{P}$   &  $\Lambda$ & $E_B$ & $\Lambda$ & $E_B$ & $\Lambda$ & $E_B$ \\    
$1^{+}$& 0.9 & \ \ 0.9&0.8&\ \ 0.2& 2.84 & 1.5 \\
            &$1.3$ &\ \  $8.1$&1.4&\ \ 6.3& 2.90& 12.8\\
             & $1.6$ & $20.1$ &1.8&15.0& 2.95 & 23.5 \\\hline

\toprule[2pt]
\end{tabular}
\end{center}

\end{table}

The $D^*\bar{D}_s+D\bar{D}^*_s$ bound state with $G_{U/V}=+$ can be related to the $Z_{cs}(3895)$ state observed at BESIII. In the literature, it is suggested as the $U/V$-spin partner of the $Z_c(3900)$ observed before the $Z_{cs}(3985)$~\cite{Meng:2020ihj}.  It should be reminded that  light meson exchanges  have considerable contribution to  binding of $D\bar{D}^*$ system.  In our previous study, the calculation suggests that the $J/\psi$ exchange is essential to reproduce the $Z_c(3900)$~\cite{He:2015mja,Ding:2020dio}.  As shown in Table~\ref{Tab: flavor factor},  only $J/\psi$ meson involves in $D^{(*)}{\bar{D}}^{(*)}_{s}$ interaction.  Hence, the $J/\psi$ exchange is essential to reproduce both $Z_{cs}(3985)$ and $Z_c(3900)$. 
In Table~\ref{Tab:  comparisons1}, we present the binding energies of the $D^{(*)}\bar{D}^{(*)}$ states with only $J/\psi$ exchange and of the $D^{(*)}{\bar{D}}^{(*)}_{s}$ systems. Generally speaking, there is no significant difference between two cases as expected, which means it is reasonable to consider $D^{(*)}{\bar{D}}^{(*)}_{s}$ state as  strange partner of $D^{(*)}{\bar{D}}^{(*)}$ state in the SU(3)$_F$ symmetry.  Moreover, we can expect that the binding of the  hidden charmed strange system $D^*{\bar{D}}_{s}+D\bar{D}^*_s$ is loosely than the hidden charmed system $D^{(*)}{\bar{D}}^{(*)}$ because more light exchanges are allowed for the latter state.

\renewcommand\tabcolsep{0.29cm}
\renewcommand{\arraystretch}{1.2}
\begin{table}[hbtp!]
\begin{center}
\caption{The binding energies of some bound states with selected value of cutoff $\Lambda$. The result for the $D^{(*)}\bar{{D}_{s}}^{(*)}$ systems are listed in the $2-4^{th}$ columns. The results for the $D^{(*)}\bar{D}^{(*)}$ only with  $J/\psi$ exchange  are listed in $5-7^{th}$ columns. The cutoff $\Lambda$ and binding energy $E_B$ are in the units of GeV and MeV.
\label{Tab:  comparisons1}
\label{diagrams}}
	\begin{tabular}{lcccccc}\bottomrule[2pt]\hline
 $J^{PG}$&\multicolumn{3}{c}{$Z_{cs}$}&\multicolumn{3}{c}{$Z_{c}$ only with $J/\psi$} \\\cmidrule(lr){1-1}\cmidrule(lr){2-4}\cmidrule(lr){5-7}
& system  &  $\Lambda$ & $E_B$& system & $\Lambda$ & $E_B$  \\
$0^{+} $ &$D\bar{D}_{s}$& $ 4.6$ &  $ 5.4$&$D\bar{D}$& $4.6$ &$0.3$ \\
$1^{++}$& $D^*{\bar{D}}_{s}+D\bar{D}^*_s$& $3.2$ &  $ 5.0$& $D{\bar{D}}^{*}$& $3.2$ &  $0.8$  \\
$0^{+}$&${D}^{*}{\bar{D}}^{*}_{s}$& $4.4$ &  $ 4.8$ &${D}^{*}{\bar{D}}^{*}$&  $4.4$ &$0.6$ \\
$1^{+}$&${D}^{*}{\bar{D}}^{*}_{s}$& $4.4$ &  $ 5.5$ &${D}^{*}{\bar{D}}^{*}$&  $4.4$ &$0.9$ \\
$2^{+}$&${D}^{*}{\bar{D}}^{*}_{s}$& $4.4$ &  $ 6.1$&${D}^{*}{\bar{D}}^{*}$&  $4.4$ &$1.1$  \\
\hline
\toprule[2pt]
\end{tabular}
\end{center}

\end{table}

In Table~\ref{Tab: DDA bound state} we also present results for the systems ${D_{s}}{\bar{D}}^*_{s}$/${B_{s}}{\bar{B}}^*_{s}$ with different $C$ parities.  One can find that the bound states are produced from both ${D_{s}}{\bar{D}}^*_{s}$ and ${B_{s}}{\bar{B}}^*_{s}$ interactions with two $C$ parties at small cutoffs. The two bound states of the ${D_{s}}{\bar{D}}^*_{s}$ system appears at cutoff of about 1.5~GeV, and the ${B_{s}}{\bar{B}}^*_{s}$ system is bound at a smaller cutoff of about 1~GeV. For doubly heavy systems  ${D_{s}}{{D}}^*_{s}$/${B_{s}}{{B}}^*_{s}$,  the spin parity is $1^+$.  A bound state can be found at a cutoff of about 3.4~GeV from the $D_sD^*_s$ interaction and a bound state can be produced at a cutoff of about 2.8~GeV from the $B_sB^*_s$ interaction.

In Table~\ref{Tab: DADA bound state}, the binding energies of the bound states with two vector mesons, $D^*_{(s)}\bar{D}^*_s/B^*_{(s)}\bar{B}^*_s$, and $D^*_{(s)}{D}^*_s/B^*_{(s)}{B}^*_s$, are presented.  Three spin parities $0^+$, $1^+$, and $2^+$ can be produced from S-wave interaction of two vector mesons.   For the $D^*_s\bar{D}^*_s$ system, a bound state  with $1^+$ appears at cutoff  of 1.7~GeV.  The $D^*_s\bar{D}^*_s$ bound states with $0^+$ and $2^+$ are also produced with a relatively small cutoff of 1.4 and 2.0~GeV, respectively, which is consistent with  the prediction of existence  of a $0^{++}$ molecular state from the $D^*_s\bar{D}^*_s$ interaction~\cite{Liu:2009ei,Albuquerque:2009ak,Zhang:2009st,Wang:2014gwa,Chen:2015fdn}.  The bound states are produced from the  $D^*_s\bar{D}^*_s$ interaction with three spins, but cutoffs larger than 4~GeV are required. In the bottom sector, the $B^*\bar{B}^*_s$ interaction is still attractive for three spins, but not strong enough to produce bound state at a cutoff smaller than 5 GeV. Three bound states can be found in the $B^*_s\bar{B}^*_s$ interaction for three spins at small cutoff as the $D^*_s\bar{D}^*_s$ interactions. 
The cutoffs to produce ${D}_{s}^{(*)}{\bar{D}}^{(*)}_{s}$ and ${{B}_s}^{(*)}{\bar{B}}^{(*)}_{s}$ states are much smaller than these for $D^{(*)}{\bar{D}}^{(*)}_{s}$ and $B^{(*)}{\bar{B}}^{(*)}_{s}$ states because exchanges of light meson $\eta$ and $\phi$  take part in former interactions. 

\renewcommand\tabcolsep{0.23cm}
\renewcommand{\arraystretch}{1.2}
\begin{table}[hbtp!]
\begin{center}
\caption{The binding energies of the bound states from the  interactions $D^*_{(s)}\bar{D}^*_s/B^*_{(s)}\bar{B}^*_s$, and $D^*_{(s)}{D}^*_s/B^*_{(s)}{B}^*_s$ with some selected values of cutoff $\Lambda$.  Other notations are the same as Table~\ref{Tab: DD bound state}.
\label{Tab: DADA bound state}
\label{diagrams}}
\begin{tabular}{ccccccccccc}\bottomrule[2pt]\hline
     $$ & \multicolumn{2}{c}{$D^*\bar{D}^{*}_{s}$}& \multicolumn{2}{c}{$D_{s}^*\bar{D}^{*}_{s}$} &\multicolumn{2}{c}{$D^*D_{s}^*$}&\multicolumn{2}{c}{$D_{s}^*D_{s}^*$}\\\cmidrule(lr){2-3}\cmidrule(lr){4-5}\cmidrule(lr){6-7}\cmidrule(lr){8-9}
     $J^{P}$   &  $\Lambda$ & $E_B$ & $\Lambda$ & $E_B$ & $\Lambda$ & $E_B$ &  $\Lambda$ & $E_B$ \\
$0^{+}$ &$4.2$  &\ \  $1.3$&$1.3$  &\ \  $0.9$&$\tdots$& $\tdots$&$\tdots$& $\tdots$ \\
             &$4.5$  &\ \  $7.8$& $1.6$  & \ \ $7.8$  &$\tdots$& $\tdots$&$\tdots$& $\tdots$\\
             & $4.7$  & $15.4$&$1.8$  & $16.6$&$\tdots$& $\tdots$&$\tdots$& $\tdots$\\\cmidrule(lr){2-3}\cmidrule(lr){4-5}\cmidrule(lr){6-7}\cmidrule(lr){8-9}
     $J^{P}$   &  $\Lambda$ & $E_B$ & $\Lambda$ & $E_B$ & $\Lambda$ & $E_B$ &  $\Lambda$ & $E_B$ \\
$1^{+}$& 4.2 &\ \  1.7& 1.4 &\ \  0.6 &2.95 &\ \ 0.5& $3.00$& \ \ $0.4$ \\
              & 4.5 &\ \  8.4&  1.8 &\ \  7.7&3.03&\ \  8.2& $3.20$& $11.0$ \\
            & 4.8 & 22.6&2.1 & 18.5 &3.07 & 23.2 &$3.25$& $18.8$ \\\cmidrule(lr){2-3}\cmidrule(lr){4-5}\cmidrule(lr){6-7}\cmidrule(lr){8-9}
     $J^{P}$   &  $\Lambda$ & $E_B$ & $\Lambda$ & $E_B$ & $\Lambda$ & $E_B$ &  $\Lambda$ & $E_B$ \\
$2^{+}$ & 4.1 &\ \  0.8& 2.0 &\ \  1.1 &$3.21$&\ \  $0.4$& 3.00 &\ \  0.9 \\
             & 4.5 &\ \  9.2& 2.8 & \ \ 8.4&$3.23$& $10.2$& 3.10 & 11.2 \\
             &4.7& 18.1 &3.2 & 16.1&$3.24$& $25.7$&3.15& 21.6\\\hline
     $$ & \multicolumn{2}{c}{$B^*\bar{B}^{*}_{s}$}& \multicolumn{2}{c}{$B_{s}^*\bar{B}^{*}_{s}$}&\multicolumn{2}{c}{$B^*B_{s}^*$}&\multicolumn{2}{c}{$B_{s}^*B_{s}^*$} \\\cmidrule(lr){2-3}\cmidrule(lr){4-5}\cmidrule(lr){6-7}\cmidrule(lr){8-9}
     $J^{P}$   &  $\Lambda$ & $E_B$ & $\Lambda$ & $E_B$ & $\Lambda$ & $E_B$ &  $\Lambda$ & $E_B$ \\
$0^{+}$  & $\tdots$  &$\tdots$&$0.8$  & \ \ $0.7$ & $3.9$&\ \  $1.4$&$\tdots$&$\tdots$ \\
             & $\tdots$ & $\tdots$& $1.2$  &\ \  $8.2$ & $4.2$  &\ \  $8.7$&$\tdots$&$\tdots$\\
              & $\tdots$  & $\tdots$& $1.5$  & $20.5$ & $4.4$ & $20.3$ &$\tdots$& $\tdots$\\\cmidrule(lr){2-3}\cmidrule(lr){4-5}\cmidrule(lr){6-7}\cmidrule(lr){8-9}
     $J^{P}$   &  $\Lambda$ & $E_B$ & $\Lambda$ & $E_B$ & $\Lambda$ & $E_B$ &  $\Lambda$ & $E_B$ \\
$1^{+}$& $\tdots$ &$\tdots$&0.8 &\ \  0.4 & 1.2  &\ \  0.7&1.8&\ \  0.7 \\
            &$\tdots$& $\tdots$& 1.2& \ \ 5.7& 1.7 & \ \ 8.8& 2.3&\ \  6.8  \\
             & $\tdots$ & $\tdots$&1.6 & 18.3 & 2.0 &19.8 &2.6& 14.3\\\cmidrule(lr){2-3}\cmidrule(lr){4-5}\cmidrule(lr){6-7}\cmidrule(lr){8-9}
     $J^{P}$   &  $\Lambda$ & $E_B$ & $\Lambda$ & $E_B$ & $\Lambda$ & $E_B$ &  $\Lambda$ & $E_B$ \\
$2^{+}$ &$\tdots$&$\tdots$& 1.0 & \ \ 0.7&$2.0$& \ \ $0.6$ & $2.1$ &\ \ $0.2$ \\
             & $\tdots$& $\tdots$& 1.8 & \ \ 8.1 & $2.7$ &\ \ $8.7$ &$2.5$ & $11.5$  \\
             &$\tdots$&$\tdots$&2.2 &  13.8 &$2.9$& $15.3$ &$2.6$& $19.2$\\\hline
\toprule[2pt]
\end{tabular}
\end{center}

\end{table}

In the doubly charmed sector, no bound state is found from the $D^*D^*_s$ and $D^*_sD^*_s$ interactions with spin parity $0^+$ while bound states can be produced at cutoffs of about 3~GeV for spin parities $1^+$ and $2^+$.  In the doubly bottom sector, the interactions become more attractive, one can find small cutoffs are required compared with these in the charmed sector, and a bound state can be found from the $B^*B^*_s$ interaction with $0^+$.

 In our calculation we adopt the widely used assumption that  no $s\bar{s}$ component in $\sigma$ meson. Hence, it can not be exchanged between the mesons considered in the current work. In Ref.~\cite{Yan:2021tcp} where the $J/\psi$ exchange was not included, the $\sigma$ exchange is proposed to play the most important factor to form a molecular state to interpret the  ${Z_{cs}}(3985)^-$.  As a complement, the influence of $\sigma$ exchange in our calculation is listed in Table~\ref{Tab:  comparisons2}. Here, the Lagrangians in Eq.~(\ref{Eq: L}) for the vertices ${\cal P}{\cal P}\sigma$ and ${\cal P}^*{\cal P}^*\sigma$ are also applied to the heavy-strange meson. The binding energies of bound states with  selected value of cutoff $\Lambda$ are presented.  One can find that there is no significant difference between  cases with and without $\sigma$ exchange. It suggests that if  we include the $J/\psi$ exchange, the contribution from the $\sigma$ exchange may be smeared in our theoretical frame.
\renewcommand\tabcolsep{0.36cm}
\renewcommand{\arraystretch}{1.1}
\begin{table}[hbtp!]
\begin{center}
\caption{The binding energies of some bound states with some selected values of cutoff $\Lambda$. The results  with  $\sigma$ exchange are listed $3-4^{rd}$ columns. The result without  $\sigma$ exchange are listed $5-6^{th}$ columns. The cutoff $\Lambda$ and binding energy $E_B$ are in the units of GeV and MeV.}
\label{Tab:  comparisons2}
\label{diagrams}
	\begin{tabular}{clrrrr}\bottomrule[2pt]\hline
system &$J^{PC/G}$&\multicolumn{2}{c}{with $\sigma$ }&\multicolumn{2}{c}{without $\sigma$} \\\cmidrule(lr){1-2}\cmidrule(lr){3-4}\cmidrule(lr){5-6}
 &  &  $\Lambda$ & $E_B$ & $\Lambda$ & $E_B$   \\
$D\bar{D}_s$&$0^{+} $&$4.4$ &$6.7$ & $ 4.4$ &  $ 1.1$ \\
$D^*{\bar{D}}_{s}+D\bar{D}^{*}_{s}$&$ 1^{++}$& $3.1$ &  $5.9$ & $3.1$ &  $ 1.4$ \\
$ {D}^{*}{\bar{D}}^{*}_{s}$&$ 0^{+}$& $4.2$ &$6.1$ & $4.2$ &  $ 1.3$ \\
${D}^{*}{\bar{D}}^{*}_{s}$&$1^{+}$& $4.2$ &$6.8$ & $4.2$ &  $ 1.7$ \\
${D}^{*}{\bar{D}}^{*}_{s}$&$ 2^{+}$& $4.1$ &$5.1$ & $4.1$ &  $ 0.8$ \\
\cmidrule(lr){1-2}\cmidrule(lr){3-4}\cmidrule(lr){5-6}
$D{D_s}$&$ 0^{+} $&$\tdots$ &$\tdots$ &$\tdots$&$\tdots$ \\
$D^*{{D}}_{s}+ DD_s^*$&$ 1^{++}$ & $\tdots$ & $\tdots$& $\tdots$ & $\tdots$\\
$ {D}^{*}{D_s^*}$&$ 0^{+}$& $\tdots$ &  $\tdots$ & $\tdots$ &  $\tdots$ \\
${D}^{*}{D_s^*}$&$ 1^{+}$ & $2.95$ &  $ 1.2$&  $2.95$ &$0.5$ \\
${D}^{*}{D_s^*}$&$2^{+}$& $3.21$ &  $ 1.5$ & $3.21$ &$0.4$ \\
   \cmidrule(lr){1-2}\cmidrule(lr){3-4}\cmidrule(lr){5-6}
$D_s\bar{D}_s$&$ 0^{+} $&$1.6$ &$1.1$ & $ 1.6$ &  $ 0.3$ \\
$D_s{\bar{D}}^{*}_{s}$&$ 1^{++}$& $1.4$ &  $1.9$ & $1.4$ &  $ 0.8$ \\
$ {D}^{*}_s{\bar{D}}^{*}_{s}$&$0^{+}$& $1.3$ &$1.9$ & $1.3$ &  $ 0.9$ \\
${D}^{*}_s{\bar{D}}^{*}_{s}$&$1^{+}$& $1.4$ &$1.6$ & $1.4$ &  $ 0.6$ \\
${D}^{*}_s{\bar{D}}^{*}_{s}$&$2^{+}$& $2.0$ &$2.6$ & $2.0$ &  $ 1.1$ \\
\cmidrule(lr){1-2}\cmidrule(lr){3-4}\cmidrule(lr){5-6}
$D_s{D_s}$&$ 0^{+} $&$\tdots$ &$\tdots$ &$\tdots$&$\tdots$ \\
$ D_s{D_s^*}$&$ 1^{+}$&$3.41$ &$6.1$  & $3.41$ &$1.2$ \\
$ {D}^{*}_s{D_s^*}$&$ 0^{+}$&$\tdots$ &$\tdots$  & $\tdots$ &$\tdots$ \\
${D}^{*}_s{D_s^*}$&$ 1^{+}$&  $3.0$ &$0.4$ & $3.0$ &  $ 0.4$ \\
${D}^{*}_s{D_s^*}$&$ 2^{+}$& $3.0$ &$1.3$ & $3.0$ &  $ 0.9$ \\
\hline
\toprule[2pt]
\end{tabular}
\end{center}

\end{table}

\subsection{Numerical results with coupled-channel calculation}
Now, we consider the coupled-channel effect between the systems considered in the current work. The results for hidden charmed systems and doubly charmed systems  are listed in Tables~\ref{Tab: CC1} and \ref{Tab: CC2}, respectively.  Only the systems with  the same quark constitutes and spin parity can be coupled. With couplings, the position $z$ of states above the lowest threshold  involved will acquire an imaginary part, which corresponds to the width as $\Gamma=-2 {\rm Im} z$. To compare with the single-channel results, in the Tables we present the position as  $M_{th}-z$  instead of the position  $z$ of the pole, with the $M_{th}$ being the nearest threshold.

\renewcommand\tabcolsep{0.17cm}
\renewcommand{\arraystretch}{1.2}
\begin{table}[h!]
\begin{center}
\caption{The $M_{th}-z$ of the poles from the hidden-heavy coupled-channel interaction. The cutoff $\Lambda$ and position of the pole $z$ are in units of GeV and MeV, respectively. }
\label{Tab: CC1}
\begin{tabular}{ccccccc}\bottomrule[2pt]\hline
\multicolumn{3}{c}{$0^{+}$}&\multicolumn{4}{c}{$1^{+}$}\\\cmidrule(lr){1-3}\cmidrule(lr){4-7}
 $\Lambda$& \multicolumn{2}{c}{$M_{th}-z$} &
  $\Lambda$ &\multicolumn{3}{c}{$M_{th}-z$} \\\cmidrule(lr){2-3}\cmidrule(lr){5-7}
  & $D{\bar{D}}_{s}$ & $D^{*}{\bar{D}}^{*}_{s}$&
  &$D^{*}{\bar{D}}_{s}$ &$D{\bar{D}}^{*}_{s}$&$D^{*}{\bar{D}}^{*}_{s}$\\
4.2  &$1.9$ &\ \ $1.7+4.2i$&3.7&0.9& $\tdots$& $\tdots$\\
 4.3  &$6.0$ &\ \ $3.8+7.0i$&3.9&10.0&$1.0+0.87i$& $\tdots$\\
  4.4  &$13.4$ &$5.8+11.1i$&4.0&19.2&$2.8+0.00i$&$0.2+0.04i$\\
   &$$ &$$&4.1&35.3&$4.1+0.00i$&$0.7+0.13i$\\
\multicolumn{3}{c}{$0^{+}$}&\multicolumn{3}{c}{$1^{+-}$}\\\cmidrule(lr){1-3}\cmidrule(lr){4-6}
 $\Lambda$& \multicolumn{2}{c}{$M_{th}-z$} &
  $\Lambda$ &\multicolumn{2}{c}{$M_{th}-z$} \\\cmidrule(lr){2-3}\cmidrule(lr){5-6}
& $D_s{\bar{D}}_{s}$ & $D_s^{*}{\bar{D}}^{*}_{s}$&
  &$D_s^{*}{\bar{D}}_{s}$ &$D_s^{*}{\bar{D}}^{*}_{s}$\\
1.6  &$0.3$ &\ $7.8+0.11i$&1.4&0.8&$0.6+0.00i$&\\
 1.7  &$1.0$ &$11.7+0.21i$&1.5&2.5&$1.7+0.01i$&\\
  1.8  &$2.2$ &$16.6+0.37i$&1.6&5.4&$3.2+0.01i$&\\

\multicolumn{3}{c}{$0^{+}$}&\multicolumn{4}{c}{$1^{+}$}\\\cmidrule(lr){1-3}\cmidrule(lr){4-7}
 $\Lambda$& \multicolumn{2}{c}{$M_{th}-z$} &
  $\Lambda$ &\multicolumn{3}{c}{$M_{th}-z$} \\\cmidrule(lr){2-3}\cmidrule(lr){5-7}
 & $B{\bar{B}}_{s}$ & $B^{*}{\bar{B}}^{*}_{s}$&
 &$B^{*}{\bar{B}}_{s}$ &$B{\bar{B}}^{*}_{s}$ &$B^{*}{\bar{B}}^{*}_{s}$\\
$\tdots$& $\tdots$&$\tdots$&$\tdots$&$\tdots$&$\tdots$&$\tdots$\\
 $\tdots$& $\tdots$&$\tdots$&$\tdots$&$\tdots$&$\tdots$&$\tdots$\\
  $\tdots$& $\tdots$&$\tdots$&$\tdots$&$\tdots$&$\tdots$&$\tdots$\\
\multicolumn{3}{c}{$0^{+}$}&\multicolumn{4}{c}{$1^{+}$}\\\cmidrule(lr){1-3}\cmidrule(lr){4-7}
 $\Lambda$& \multicolumn{2}{c}{$M_{th}-z$} &
  $\Lambda$ &\multicolumn{2}{c}{$M_{th}-z$} \\\cmidrule(lr){2-3}\cmidrule(lr){5-6}
 & $B_s{\bar{B}}_{s}$ & $B_s^{*}{\bar{B}}^{*}_{s}$&
  &$B_s^{*}{\bar{B}}_{s}$ &$B_s^{*}{\bar{B}}^{*}_{s}$\\
1.0  &$0.5$ &$3.4+0.00i$&1.0&1.8&$2.2+0.01i$&\\
 1.1  &$2.5$ &$5.4+0.00i$&1.1&3.3&$3.7+0.02i$&\\
  1.2  &$3.8$ &$8.2+0.00i$&1.2&5.3&$5.7+0.02i$&\\\hline
\toprule[2pt]
\end{tabular}
\end{center}
\end{table}

In the first part of Table~\ref{Tab: CC1}, we present the results for $D^{(*)}\bar{D}^{(*)}_s$ interactions. For the spin parity $0^+$, only interactions $D\bar{D}_s$ and $D^*\bar{D}_s^*$  involve in the hidden charm strange system. In Table~\ref{Tab: DD bound state} and Table~\ref{Tab: DADA bound state}, bound states are found at cutoff of 4.4~GeV and 4.2~GeV from the interactions $D\bar{D}_s$ and $D^*\bar{D}_s^*$, respectively. After coupling effect is included, two poles are found near the $D\bar{D}_s$ and $D^*\bar{D}_s^*$ thresholds at about 4.2~GeV. In the bottom sector,  as in the single-channel calculation, no bound state is produced from the $B\bar{B}_s-B^*\bar{B}_s^*$ interaction in the range of cutoff considered. 
For hidden charm hidden-strange system $D_s\bar{D}_s-D^*_s\bar{D}^*_s$ with $0^+$, two poles are produced near two thresholds at a cutoff of about 1.5~GeV, and the variation of the position of lower pole is much slower  than the higher one, which is consistent with the results in the single-channel calculation.  The result for the system $B_s\bar{B}_s-B^*_s\bar{B}^*_s$ with $0^+$ is analogous but with smaller cutoff. 

For hidden charmed strange system with $1^+$, three  channels $D^*\bar{D}_s$, $D\bar{D}^*_s$, and $D^*\bar{D}_s^*$ involve. In the single-channel calculation,  we study the former two channels by constructing a U/V spin wave function under the SU(3)$_F$ symmetry. There,  the masses of $D_s$ and $D$, as well as $D^*_s$ and $D^*$ mesons,  will be chosen as their average values. However, these masses are different in experiment, which leads to larger violation of symmetry than the $C$ parity in Eq.~(\ref{Eq: wf3}) where the difference between the masses of $D^{(*)}_s$ mesons with different charges is very small in experiment. Here we take wave functions in  Eq.~(\ref{Eq: wf2}) to perform the coupled-channel calculation to discuss the effect of violation of symmetry, and results for $D^*\bar{D}_s-D\bar{D}^*_s-D^*\bar{D}^*_s$ system with $1^+$ are given in Table~\ref{Tab: CC1}.  Generally speaking, the results with such treatment are consistent with single-channel calculation. Poles are produced near the $D^*\bar{D}_s$ and $D\bar{D}^*_s$ thresholds but with a little larger cutoff, which reflects  effect of violation of the SU(3)$_F$ symmetry. The pole near $D^*\bar{D}_s^*$ threshold appears at a cutoff of about 4~GeV, which is analogous to the single-channel result. 

For the hidden charm hidden strange systems, the wave functions can be constructed with fixed $C$ parity, which can be well defined. Since the $D_s^*\bar{D}_s^*$ has a $C=-$ parity, we only consider $D^*_s\bar{D}_s-D^*_s\bar{D}^*_s$ system with $1^{+-}$. Two poles are reproduced near $D^*_s\bar{D}_s$ and $D^*_s\bar{D}^*_s$ thresholds  at a cutoff of about 1.4~GeV.  The above results suggest that inclusion of  couplings between the channels considered in the current work affects the single-channel results very small. And  violation of SU(3)$_F$ symmetry  provides visible effect, but the conclusion is unchanged.

In Table~\ref{Tab: CC2}, the results for  doubly charmed and doubly-bottomed systems with coupled-channel calculation are presented.  No pole is found from all systems considered with $0^+$, which is consistent with the single-channel results. For strange system $DD_s^*-D^*D_s-D^*D_s^*$,  poles are found near  the $D^*D_s$ and $DD^*_s$ thresholds  with spin parity $1^+$ at cutoff of about 2.2 and 2.3~GeV, respectively. In doubly bottom sector, the poles are found at cutoffs of about 1.4 and 1.2~GeV, respectively. As in the hidden heavy sector (see Table~\ref{Tab: CC1}), such results are different from the single-channel results with $U/V$ spins quantitively, but the conclusion is analogous qualitatively.  The poles near the $D^*D^*_s$ and $B^*B^*_s$ thresholds appear at cutoffs of    2.5 and 1.2~GeV, which is similar to the values in the single-channel calculation.  For the hidden-strange system, the poles are found near two thresholds of $D^*_sD_s-D^*_sD^*_s$ system at cutoffs of 2.3~GeV and 3.3~GeV, and two thresholds of $B^*_sB_s-B^*_sB^*_s$ system at cutoffs of  1.2 and 2.6~GeV, respectively.

\renewcommand\tabcolsep{0.22cm}
\renewcommand{\arraystretch}{1.1}
\begin{table}[h!]
\begin{center}
\caption{ The $M_{th}-z$ of the poles from the doubly heavy coupled-channel interaction. The cutoff $\Lambda$ and $M_{th}-z$ are in units of GeV and MeV, respectively.}
\label{Tab: CC2}
\begin{tabular}{ccccccccc}\bottomrule[2pt]\hline
\multicolumn{3}{c}{$0^{+}$}&\multicolumn{4}{c}{$1^{+}$}\\\cmidrule(lr){1-3}\cmidrule(lr){4-7}
 $\Lambda$& \multicolumn{2}{c}{$M_{th}-z$} &
  $\Lambda$ &\multicolumn{3}{c}{$M_{th}-z$} \\\cmidrule(lr){2-3}\cmidrule(lr){5-7}
& $D{{D}}_{s}$ & $D^{*}{{D}}^{*}_{s}$&
 &$D^{*}{{D}}_{s}$  &$D{{D}}^{*}_{s}$&$D^{*}{{D}}^{*}_{s}$\\
 $\tdots$&$\tdots$& $\tdots$&2.20&\ \ 1.6&$\tdots$&$\tdots$\\
  $\tdots$&$\tdots$& $\tdots$&2.33&16.7&$0.6+1.7i$&$\tdots$\\
    $\tdots$&$\tdots$& $\tdots$&2.40&32.5&$5.2+0.0i$&$1.5+3.9i$\\
   $\tdots$ &$\tdots$& $\tdots$&2.50&$\tdots$&20.4+0.0i&$4.4+5.4i$\\
\multicolumn{3}{c}{$0^{+}$}&\multicolumn{3}{c}{$1^{+}$}\\\cmidrule(lr){1-3}\cmidrule(lr){4-6}
 $\Lambda$& \multicolumn{2}{c}{$M_{th}-z$} &
  $\Lambda$ &\multicolumn{2}{c}{$M_{th}-z$} \\\cmidrule(lr){2-3}\cmidrule(lr){5-6}
 & $D_s{{D}}_{s}$ & $D_s^{*}{{D}}^{*}_{s}$&
 &$D_s^{*}{{D}}_{s}$ &$D_s^{*}{{D}}^{*}_{s}$\\
  $\tdots$&$\tdots$& $\tdots$&$2.30$ &$\tdots$&$0.2+1.8i$\\
   $\tdots$&$\tdots$& $\tdots$&$2.70$ &$\tdots$&$35.0+6.9i$\\
      $\tdots$&$\tdots$& $\tdots$&$3.31$ &$0.1$&$\tdots$\\
            $\tdots$&$\tdots$& $\tdots$&$3.40$ &$21.9$&$\tdots$\\

\multicolumn{3}{c}{$0^{+}$}&\multicolumn{4}{c}{$1^{+}$}\\\cmidrule(lr){1-3}\cmidrule(lr){4-7}
 $\Lambda$& \multicolumn{2}{c}{$M_{th}-z$} &
  $\Lambda$ &\multicolumn{2}{c}{$M_{th}-z$} \\\cmidrule(lr){2-3}\cmidrule(lr){5-7}
 & $B{{B}}_{s}$ & $B^{*}{{B}}^{*}_{s}$&
 & $B^{*}{{B}}_{s}$& $B{{B}}^{*}_{s}$&$B^{*}{{B}}^{*}_{s}$\\
 $\tdots$ &$\tdots$& $\tdots$&1.2&$\tdots$&\ \ $4.9+0.0i$&$1.7+2.2i$\\
 $\tdots$&$\tdots$& $\tdots$&1.3& $\tdots$&$7.7+0.0i$&$4.3+3.0i$\\
$\tdots$&$\tdots$& $\tdots$&1.4& $0.1$&$11.7+0.0i$&$5.2+4.2i$\\
\multicolumn{3}{c}{$0^{+}$}&\multicolumn{3}{c}{$1^{+}$}\\\cmidrule(lr){1-3}\cmidrule(lr){4-6}
 $\Lambda$& \multicolumn{2}{c}{$M_{th}-z$} &
  $\Lambda$ &\multicolumn{2}{c}{$M_{th}-z$} \\\cmidrule(lr){2-3}\cmidrule(lr){5-6}
 & $B_s{{B}}_{s}$ & $B_s^{*}{{B}}^{*}_{s}$&
 &$B_s^{*}{{B}}_{s}$ &$B_s^{*}{{B}}^{*}_{s}$\\
 $\tdots$&$\tdots$& $\tdots$&$1.2$ &$\tdots$&$0.2+0.66i$\\
  $\tdots$ &$\tdots$& $\tdots$ &$2.1$ &$\tdots$&$36.7+7.7i$\\
$\tdots$ &$\tdots$& $\tdots$ &$2.6$ &$4.8$&$\tdots$\\
  $\tdots$&$\tdots$& $\tdots$&$2.8$ &$21.0$&$\tdots$\\\hline
\toprule[2pt]
\end{tabular}
\end{center}
\end{table}

\section{Summary and discussion} \label{Sec: Summary}
 
Inspired by the newly observed ${Z_{cs}}(3985)^-$, we study possible hidden and doubly heavy molecular states with hidden and open strangeness from interactions of $D^{(*)}{\bar{D}}^{(*)}_{s}$/$B^{(*)}{\bar{B}}^{(*)}_{s}$,  ${D}^{(*)}_{s}{\bar{D}}^{(*)}_{s}$/${{B}}^{(*)}_{s}{\bar{B}}^{(*)}_{s}$, ${D}^{(*)}{D_{s}}^{(*)}$/${B}^{(*)}{B_{s}}^{(*)}$ and ${D_{s}}^{(*)}{D_{s}}^{(*)}$/${B_{s}}^{(*)}{B_{s}}^{(*)}$ in a qBSE approach. In the interactions of these systems, the light meson exchanges, as well as the $J/\psi/\Upsilon$ exchange, are included to construct the potential to find molecular state as a pole of the scattering amplitude. 

The $D\bar{D}^*_s$ interaction is considered with the $U/V$ spins,  a bound state is produced at a cutoff of about 3~GeV with spin parity $1^+$ and $G_{U/V}=+$,  which can be related to the experimentally observed $Z_{cs}(3985)$. An obvious difference between the hidden charmed interaction and hidden charm strange interaction is that only $J/\psi$ exchange happens for the latter.  If we take only the $J/\psi$ exchange,  an explicit calculation suggests that very similar binding energies can be obtained for two systems, which suggests that the $Z_{cs}(3985)$ can be seen as a partner of $Z_c(3900)$ state with $I^G(J^{PC})=1^+(1^{+-})$ in the molecular state picture. Since light meson exchange is possible for  hidden charm state, the  $Z_c(3900)$ state is more deeply bound than the $Z_{cs}(3985)$ in molecular state picture. The interaction of  $D^*\bar{D}^*_s$ and its bottom partner are also attractive, but much weaker than $D\bar{D}^*_s$ interaction, which makes it difficult to produce a bound state at a cutoff required to reproduce the the $Z_{cs}(3985)$.

Besides above states, the calculation favors  the existence of hidden-heavy states $D_s\bar{D}_s/B_s\bar{B}_s$  with $0^+$, $D_s\bar{D}^*_s/B_s\bar{B}^*_s$ with $1^{+\pm}$, $D^*_s\bar{D}^*_s/B^*_s\bar{B}^*_s$ with $0^+$, $1^+$, and $2^+$.  The experimental  search for the $D^*_s\bar{D}^*_s/B^*_s\bar{B}^*_s$ molecular states is strongly suggested by our results.  In the doubly heavy sector, the bound states can be found from the interactions $DD^*_s/BB^*_s$ with $1^+$ and $G'=-$, $D_s\bar{D}_s^*/B_s\bar{B}_s^*$ with $1^+$, $D^*D^*_s/B^*B^*_s$ with $1^+$ and $2^+$, and $D^*_sD^*_s/B^*_sB^*_s$ with $1^+$ and $2^+$. Some other interactions are also found attractive, but may be not strong enough to produce a bound state.  

In the calculation, we discuss the roles of the meson exchanges. The 
$J/\psi/\Upsilon$ exchange is found very important to form a molecular state. The contribution of such exchange overwhelms the contributions from light meson exchanges, as well as the $\sigma$ exchange in our approach, which need future studies. In Ref.~\cite{He:2015mja}, an analysis also suggests that the $J/\psi/\Upsilon$ exchange is equivalent to the contact interaction in other approach~\cite{Aceti:2014uea}.  The current results seem to show the importance of  short-range interaction in  formation of a molecular states, which is consistent with the conclusion in our previous work~\cite{He:2015mja,Ding:2020dio} and  other approach~\cite{Aceti:2014uea}, where the short-range interaction is found essential to reproduce the  $Z_c(3900)$.  We also perform a coupled-channel calculation to check the effect of coupling between different interactions. The results suggest that the conclusion is almost unchanged compared with the single-channel calculation. Of course, in the current work, only channels composed of  open heavy flavor mesons are considered, further studies are required to give more precise prediction.

\vskip 10pt

\noindent {\bf Acknowledgement} We thank Dr. Lu Meng for helpful comments. This project is supported by the National Natural Science
Foundation of China with Grants No. 11675228.

\end{document}